# Ultrasensitive, high-dynamic-range and broadband strain sensing by time-of-flight detection with femtosecond-laser frequency combs

Xing Lu[1]*, Shuangyou Zhang[2]*, Xing Chen[3], Dohyeon Kwon[2], Chan-Gi Jeon[2], Zhigang Zhang[3], Jungwon Kim[2]★ and Kebin Shi[1,4]★

[1]State Key Laboratory for Mesoscopic Physics, Collaborative Innovation Center of Quantum Matter, School of Physics, Peking University, Beijing 100871, China
[2]School of Mechanical and Aerospace Engineering, Korea Advanced Institute of Science and Technology (KAIST), Daejeon 34141, South Korea
[3]State Key Laboratory of Advanced Optical Communication System and Networks, School of Electronics Engineering and Computer Science, Peking University, Beijing 100871, China
[4]Collaborative Innovation Center of Extreme Optics, Shanxi University, Taiyuan, Shanxi 030006, China
*These authors contributed equally to this work
★Corresponding authors email: kebinshi@pku.edu.cn, jungwon.kim@kaist.ac.kr

**Ultrahigh-resolution optical strain sensors provide powerful tools in various scientific and engineering fields, ranging from long-baseline interferometers to civil and aerospace industries. Here we demonstrate an ultrahigh-resolution fibre strain sensing method by directly detecting the time-of-flight (TOF) change of the optical pulse train generated from a free-running passively mode-locked laser (MLL) frequency comb. We achieved a local strain resolution of 18 p$\varepsilon$/Hz$^{1/2}$ and 1.9 p$\varepsilon$/Hz$^{1/2}$ at 1 Hz and 3 kHz, respectively, with large dynamic range of >154 dB at 3 kHz. For remote-point sensing at 1-km distance, 80 p$\varepsilon$/Hz$^{1/2}$ (at 1 Hz) and 2.2 p$\varepsilon$/Hz$^{1/2}$ (at 3 kHz) resolution is demonstrated. While attaining both ultrahigh resolution and large dynamic range, the demonstrated method can be readily extended for multiple-point sensing as well by taking advantage of the broad optical comb spectra. These advantages may allow various applications of this sensor in geophysical science, structural health monitoring, and underwater science.**

Optical strain sensors are playing an increasingly important role in various scientific and engineering applications[1-10]. Ultrahigh-performance strain sensors have enabled new scientific facilities and instruments such as long-baseline interferometers for gravitational-wave detection[1,2] and geophysical observations[3,4]. Fibre-optic strain sensors have been applied to a wide range of engineering applications such as seismic monitoring[5], structural health monitoring[6,7], nanotechnology[8], and civil and aerospace industries[9,10], due to their advantages of robustness, flexibility, and high sensitivity. For more than three decades, various types of optical strain sensors[5,9,11-14] have been demonstrated. For example, strain sensors based on Brillouin scattering can monitor the distributed strain along the fibre link with millimeter spatial resolution and µε-scale strain resolution[13]. Fibre-Bragg-grating (FBG) strain sensors have been demonstrated a sub-nano-strain for static (e.g., <10 Hz) strain sensing[9] and sub-pico-strain for the dynamic (e.g., >10 Hz) strain sensing[11] with µε-scale measurement range. The well-known advantages such as cost-effectiveness, flexibility and insusceptibility to ambient electromagnetic noise have led to commercialization of these strain sensors. More recently, the incorporation of dual-comb techniques have led to the FBG strain sensor with a mε-scale measurement range with a 34-nε resolution[14].

Although dynamic strain sensing has already achieved sub-pico-strain resolution[11], static strain sensing is still hampered by the 1/$f$-noise from low-frequency noise in lasers and ambient perturbations. As a result, for high-resolution static strain sensing, an extra-reference (such as stabilized frequency combs[9], atomic absorption line[15,16] or an identical but strain-free sensor head[5,17,18]) is usually needed to compensate for the environmental temperature drift. Among static sensing methods, FBG-sensors interrogated by narrow-linewidth continuous-wave (CW) lasers can achieve the highest resolution. In FBG sensors, both wavelength-swept and wavelength-fixed CW lasers have been used to interrogate the FBG output[5,9,15-20]. The wavelength-sweeping scheme can offer ~10-n$\varepsilon$ static strain resolution[19], nevertheless, the necessity for frequency scanning usually hinders the data acquisition rate (less than 10 Hz). Besides, the resolution is also limited by the repeatability and nonlinearity of the spectral sweeping[5]. The sweeping-free scheme can overcome this resolution limitation and achieve sub-n$\varepsilon$ strain detection at low frequency, where the CW lasers are locked to the optical references[9,15,16,18] by the Pound-Drever-Hall (PDH) techniques[21]. With PDH techniques, unprecedented resolution of 0.55 p$\varepsilon$/Hz$^{1/2}$ at ~2 Hz has been recently achieved[9]. However, due to the laser locking and error detection process, its dynamic range and multiplexing capability are sacrificed. As a result, a tradeoff has to be made between resolution, measurement speed, dynamic range, and multiplexing capability.

In order to simultaneously achieve high-resolution measurement of static and dynamic strain with a large dynamic range, we propose and demonstrate a new class of fibre strain sensing method by detecting the change in the TOF of the strain-influenced femtosecond optical pulse trains generated from MLL frequency comb sources[22,23]. Strain sensing by time-domain interrogation has been explored since early 1990s[24], however, their performance was limited by the long pulsewidth (~ns long) and electronic timing discriminator (~10-ps resolution)[24,25]. To overcome these limitations, we employ femtosecond optical pulse trains generated from the MLL frequency comb sources and sub-femtosecond-resolution phase detection between optical pulses and microwave signals.

## Results

**Frequency comb TOF-based strain sensing concept.** Figure 1 shows the conceptual diagram. First, a strain-free optical pulse train is tightly synchronized with a microwave signal (Fig. 1**a**). After the interrogating pulse train goes through the strain sensor, the timing of the interrogating pulse is changed by the strain as shown in Fig. 1**b**. By comparing the timing between the synchronized microwave zero-crossing with the interrogating pulse positions, we can extract the strain signal. Here, the key challenge is how to realize high-resolution timing detection and synchronization between the optical pulse train and the microwave signal. For this purpose, in this work, we employ sub-femtosecond-resolution fibre-loop optical-microwave phase detectors (FLOM-PDs)[26], which have been recently used for low-noise microwave generation[27], frequency dissemination over long-distance fibre links[28], and timing of ultrafast electron diffraction apparatus[29]. When a microwave signal with an integer multiple of the MLL repetition-rate is applied to the FLOM-PD, the output error signal is proportional to the temporal position difference between the optical pulse and the microwave signal zero-crossing with only 8.9-μrad (i.e., 177-attosecond at 8 GHz) residual error in the 0.01 Hz – 100 kHz Fourier frequency range (see Supplementary Information). By using such tightly synchronized microwave phase as the strain sensing ruler, both high resolution and large dynamic range in TOF detection are possible. Due to the ultralow timing jitter property of typical free-running femtosecond Er-fibre MLLs[23,30], high-quality synchronization and high-resolution TOF detection over broad bandwidth is possible, which is useful for high-resolution dynamic sensing over broad bandwidth. Note that, however, for high-resolution static sensing, such low-jitter property is not necessarily required since the residual phase noise/timing jitter is fully suppressed by the laser-microwave synchronization. This greatly relaxes the requirement for noise performance of usable mode-locked lasers in our strain sensing systems.

**Local strain measurement results**. Figure 2a shows the schematic of the local strain sensing experiment. The optical pulse train generated from the MLL is divided into two paths by a coupler. One is used for generation of the reference timing by synchronizing a microwave signal by FLOM-PD1. The other is connected to the sensor head and used to interrogate the strain. Note that, as we use the pulse TOF change as a means to detect strain, the sensor head is simply a section of single-mode fibre (SMF) attached to the structure under test (see Methods). Then the synchronized microwave signal is used as a ruler to detect the timing difference between itself and the interrogating pulse by FLOM-PD2. More detailed information can be found in Methods.

We first measured the output voltage of FLOM-PD2 as a function of the applied strain (Fig. 3). As expected, large linear measurement range of >100 µε can be directly measured, benefiting from the use of microwave phase as the strain sensing ruler. Curve **a** of Fig. 4 shows the measured strain-noise power spectrum density (PSD) in local measurement from 0.01-Hz to 100-kHz band. The peak at 3 kHz in curve **a** is caused by the 3-kHz calibration signal, and the peaks at 60 Hz and its harmonics are caused by electric power supply pickup (indicated as dashed lines). The measured strain resolution at 1 Hz and 10 Hz is 18 $p\varepsilon/Hz^{1/2}$ and 7.7 $p\varepsilon/Hz^{1/2}$, respectively. For the dynamic strain, the resolution floor is ~1.9 $p\varepsilon/Hz^{1/2}$. Considering the measurement range of >100 µε, the linear dynamic range of the sensor is more than 154 dB at high frequency. At the very low frequency (below 1 Hz), the noise increases with $1/f$ dependence, reaching 1.7 $n\varepsilon/Hz^{1/2}$ at 10 mHz. As shown in Fig. 4, compared to the state-of-the-art local sensing resolutions achieved by ultrastable reference-locked CW lasers and FBGs (curves **c**, **e**, **g**, **h**)[5,9,16,17], this method also shows a good resolution and the performance is much better than free-running CW laser results (curves **d**, **i**)[9,16].

**Remote-point strain measurement result**. To verify the remote point sensing performance, as shown in Fig. 2**b**, we used two 1-km-long single-mode fibre links to connect the remote sensing point with the monitor station. For remote strain sensing, it is important to cancel out the thermal and acoustic variation noise induced by the delivery fibre link[31]: otherwise, the performance will be limited to merely µε-scale resolution by several-ps timing drift of the 2-km fibre link. To address this issue, we utilize wavelength division multiplexer (WDM) to split the MLL output spectrum into two parts: half of the spectrum is used as the reference, and the other half is used for strain sensing. The referenced path experiences the same environmental noise with the interrogating path, but it is strain-free so that the noise induced by the link can be compensated by using it as the measurement reference. By employing such environment noise cancellation method, the strain resolution could be improved by a factor of >100 in the 0.1 Hz - 1 kHz frequency range (see Supplementary Information). The remote sensing result is shown by curve **b** in Fig. 4. The measured strain resolution at 1 Hz and 10 Hz is 80 $p\varepsilon/Hz^{1/2}$ and 10 $p\varepsilon/Hz^{1/2}$, respectively. To our knowledge, this is the first time to measure remote sub-nano-level static strain over km-long fibre links. For the dynamic strain, the resolution floor is ~2.2 $p\varepsilon/Hz^{1/2}$, which is almost the same with that of the local sensing from 10 Hz to 100 kHz. It is observed that the resolution is slightly degraded for the low-frequency (0.01 Hz to 1 Hz), which, we believe, is limited by the different dispersion configuration and the non-common components such as EDFAs.

## Discussion

In summary, by detecting the TOF of the interrogating optical pulse train from a free-running MLL frequency comb source, a new strain sensing method is demonstrated with a resolution of sub-nano-strain for static and pico-strain for dynamic sensing. Compared to other schemes based on FBG interrogated by narrow-linewidth CW lasers, our method employs a free-running MLL as the interrogating source without laser frequency stabilization. It also shows both good strain resolution and large dynamic range, and has remote multiple-point sensing capability. If lower-frequency microwave (e.g., 1-GHz) is used, the demonstrated TOF strain sensing method can achieve m$\varepsilon$-scale measurement range at the expense of slightly degraded resolution (~10 $p\varepsilon/Hz^{1/2}$ floor). Due to the ultra-wide optical spectrum (as large as ~135 nm[32]) of MLL output, this method also possesses wavelength multiplexing capacity to realize multiple-point remote sensing along the fibre link. By directly using fibre as the sensor head, the proposed method particularly shows great potential in geophysical science (such as seismic and volcanic activity); underwater acoustics (such as oil and gas explorations and military surveillance) and large structural health monitoring (such as bridge and airplanes).

**Methods**

**Local strain sensing.** The 250-MHz optical pulse train generated by a free-running femtosecond Er-fibre mode-locked laser comb source (MenloSystems GmbH, FC1500-250-ULN) is divided into two paths by a 50:50 coupler, where each path has ~18-mW optical power. One path is used for the synchronization between the optical pulse train and a 8-GHz microwave signal which is generated from a voltage-controlled oscillator (VCO, Hittite HMC-C200). Microwave power of ~ +16 dBm is applied to each FLOM-PD. The phase error signal between the pulse train and the 8-GHz microwave signal is regulated by a home-built proportional-integral (PI) board, and then fed back to the VCO. The locking bandwidth is larger than 100 kHz. The other path is used for the strain measurement. In this work, to calibrate the strain of the sensor, we use a 40.8-m-long piezoelectric transducer (PZT)-driven fibre stretcher to generate a strain signal. Note that a significant amount of research and development has been performed on optimizing the size of fibre-type sensor heads (e.g., in fibre acoustic sensors), which can also benefit our proposed TOF strain sensing method by realizing strain monitoring size from ~10 mm to several metres[33]:

for example, a zig-zag pattern arrangement of fibre can significantly reduce the footprint of the sensor head while maintaining effective fibre length. To calibrate the strain, a 35-mV$_{rms}$, 3-kHz sine signal is applied to the fibre stretcher. Knowing the stretcher has a responsivity of 4.4 μm/V, the calibration corresponds to a strain of 0.96 nε/Hz$^{1/2}$ when taking the measurement bandwidth of 15.625 Hz into account. The pulse train with strain information is applied to the FLOM-PD2, and makes comparison with the phase of the microwave signal synchronized to the optical pulse train without strain (reference signal). The output error signal from FLOM-PD2 is analysed by an FFT spectrum analyser (Stanford Research Systems, SR760) for the high frequency (>1 Hz) and a data logger (Keysight, 34970A) for low frequency (<1 Hz).

**Remote-point strain sensing.** For the remote strain sensing, we used two 1-km-long delivery fibre links to connect the remote site and the local monitor station as shown in Fig. 2**b**. Here, one link is for the forward light (from local to remote), while the other is for the backward light (from remote to local). Note that we used separate fibre links for forward and backward lights to avoid fibre backscattering-induced degradation in relative intensity noise[34,35]. Two 100-m-long dispersion-compensating fibre (DCF, LLWBDK) spools are used for the dispersion compensation. Because the FLOM-PD detection sensitivity does not degrade much by pulse chirp, moderate level of dispersion compensation is enough[28]. To avoid the nonlinear effects in the fibre link, the optical power applied to the fibre link is limited to ~5 mW. At the remote sensing site, a fibre circulator is used to redirect the optical pulse to the monitor station. A WDM is used to split the optical spectrum into two parts ($\lambda_1$, 1550 nm~1558 nm and $\lambda_2$, 1558 nm~1566 nm) where $\lambda_1$ and $\lambda_2$ pulses are used for the strain-free and strain sensing reference, respectively. At the local monitor station, another WDM is used to split the light pulse into two parts ($\lambda_1$ and $\lambda_2$). As the $\lambda_1$ and $\lambda_2$ light experience the same noise in the forward and backward fibre links, only the strain signal in the $\lambda_2$ path is non-common while the noise induced by the delivery fibre link is common. In the demodulation process, we synchronize the 8-GHz microwave signal with the $\lambda_1$ part by FLOM-PD1 so that the microwave signal carries the common noise of the delivery fibre link. The FLOM-PD2 is employed to extract the phase difference between the synchronized microwave and the $\lambda_2$ optical pulse to get the strain information while the fibre noise is eliminated by the subtraction. Here, we also calibrate the strain by modulating the same 35-mV$_{rms}$, 3-kHz sine signal, and obtain the same amplitude strain. Therefore, we cancel out the delivery fibre noise and achieve an ultrasensitive strain resolution at remote sensing site.

**References**

1. Harry, G. M. & LIGO Scientific Collaboration Advanced LIGO: the next generation of gravitational wave detectors. *Classical Quant. Grav.* **27**, 084006 (2010).

2. Abbott, B. P. *et al.* Observation of gravitational waves from a binary black hole merger. *Phys. Rev. Lett.* **116**, 061102 (2016).

3. Berger, J. & Lovberg, R. Earth strain measurements with a laser interferometer.

*Science* **170**, 296–303 (1970).

4. Araya, A. *et al.* Iodine-stabilized Nd: YAG laser applied to a long-baseline interferometer for wideband earth strain observations. *Rev. Sci. Instrum.* **73**, 2434-2439 (2002).

5. Chen, J., Liu, Q., Fan, X. & He, Z. Ultrahigh resolution optical fiber strain sensor using dual Pound–Drever–Hall feedback loops. *Opt. Lett.* **41**, 1066-1069 (2016).

6. López-Higuera, J. M., Cobo, L .R., Incera, A. Q., & Cobo, A. Fiber Optic Sensors in Structural Health Monitoring. *J. Lightwave Technol.* **29**, 587-608 (2011).

7. Majumder, M., Gangopadhyay, T. K., Chakraborty, A. K., Dasgupta, K., & Bhattacharya, D. K. Fibre Bragg gratings in structural health monitoring - Present status and applications. *Sens. Actuators A*. **147**, 150–164 (2008).

8. Bitou, Y. High-accuracy displacement metrology and control using a dual Fabry-Perot cavity with an optical frequency comb generator. *Precis. Eng.* **33**, 187-193 (2009).

9. Gagliardi, G., Salza, M., Avino, S., Ferraro, P. & De Natale, P. Probing the Ultimate Limit of Fiber-Optic Strain Sensing. *Science* **330**, 1081-1084 (2010).

10. Marques dos Santos F. L. & Peeters, B. On the use of strain sensor technologies for strain modal analysis: Case studies in aeronautical applications. *Rev. Sci. Instrum.* **87**, 102506 (2016).

11. Chow, J. H., McClelland, D. E., Gray, M. B. & Littler, I. C. M. Demonstration of

a passive subpicostrain fiber strain sensor. *Opt. Lett.* **30**, 1923-1925 (2005).

12. Littler, I. C. M., Gray, M. B., Chow, J. H., Shaddock, D. A. & McClelland, D. E. Pico-strain multiplexed fiber optic sensor array operating down to infra-sonic frequencies. *Opt. Express* **17**, 11077-11087 (2009).

13. Song, K. Y., He, Z., & Hotate, K. Distributed strain measurement with millimeter-order spatial resolution based on Brillouin optical correlation domain analysis. *Opt. Lett.* **31**, 2526-2528 (2006).

14. Kuse, N., Ozawa, A. & Kobayashi, Y. Static FBG strain sensor with high resolution and large dynamic range by dual-comb spectroscopy. *Opt. Express* **21**, 11141-11149 (2013).

15. Arie, A., Lissak, B. & Tur, M. Static fiber-Bragg grating strain sensing using frequency-locked lasers. *J. Lightwave Technol.* **17**, 1849-1855 (1999).

16. Lam, T. T. Y. *et al.* High-resolution absolute frequency referenced fiber optic sensor for quasi-static strain sensing. *Appl. Opt.* **49**, 4029-4033 (2010).

17. Huang W, Feng S, Zhang W, et al. DFB fiber laser static strain sensor based on beat frequency interrogation with a reference fiber laser locked to a FBG resonator. *Opt. Express* **24**, 12321-12329 (2016).

18. Liu, Q. W., He, Z. Y. & Tokunaga, T. Sensing the earth crustal deformation with nano-strain resolution fiber-optic sensors. *Opt. Express* **23**, A428-A436 (2015).

19. Liu, Q. *et al.* Ultrahigh Resolution Multiplexed Fiber Bragg Grating Sensor for

Crustal Strain Monitoring. *IEEE Photonics J.* **4**, 996-1003 (2012).

20. Huang, W. Z., Zhang, W. T. & Li, F. Swept optical SSB-SC modulation technique for high-resolution large-dynamic-range static strain measurement using FBG-FP sensors. *Opt. Lett.* **40**, 1406-1409 (2015).

21. Black, E. D. An introduction to Pound–Drever–Hall laser frequency stabilization. *Am. J. Phys.* **69**, 79-87 (2001).

22. Cundiff, S. T., Ye, J., & Hall, J. L. Optical frequency synthesis based on mode-locked lasers. *Rev. Sci. Instrum.* **72**, 3749-3771 (2001).

23. Kim, J. & Song, Y. Ultralow-noise mode-locked fiber lasers and frequency combs: principles, status, and applications. *Adv. Opt. Photonics* **8**, 465-540 (2016).

24. Zimmerman, B. D., Claus, R. O., Kapp, D. A. & Murphy, K. A. Fiber-optic Sensors Using High-Resolution Optical Time Domain Instrumentation Systems. *J. Lightwave Technol.* **8**, 1273-1277 (1990).

25. Lyöri, V., Kilpelä, A., Duan, G., Mäntyniemi, A. & Kostamovaara, J. Pulsed time-of-flight radar for fiber-optic strain sensing. *Rev. Sci. Instrum.* **78**, 024705 (2007).

26. Jung, K. & Kim, J. Subfemtosecond synchronization of microwave oscillators with mode-locked Er-fiber lasers. *Opt. Lett.* **37**, 2958-2960 (2012).

27. Jung, K., Shin, J. & Kim, J. Ultralow phase noise microwave generation from

mode-locked Er-fiber lasers with subfemtosecond integrated timing jitter. *IEEE Photon. J.* **5**, 5500906 (2013).

28. Jung, K., Shin, J., Kang, J., Hunziker, S., Min, C. K. & Kim, J. Frequency comb-based microwave transfer over fiber with $7\times10^{-19}$ instability using fiber-loop optical-microwave phase detectors. *Opt. Lett.*, **39**, 1577-1580 (2014).

29. Yang, H. *et al.* 10-fs-level synchronization of photocathode laser with RF-oscillator for ultrafast electron and X-ray sources, *Sci. Rep.*, **7**, 39966 (2017).

30. Kwon, D. *et al.* Reference-free, high-resolution measurement method of timing jitter spectra of optical frequency combs. *Sci. Rep.* **7**, 40917 (2017).

31. Foreman, S. M., Holman, K. W., Hudson, D. D., Jones, D. J. & Ye, J. Remote transfer of ultrastable frequency references via fiber networks. *Rev. Sci. Instrum*. **78**, 021101 (2007).

32. Ma, D., Cai, Y., Zhou, C., Zong, W. J., Chen, L. L. & Zhang, Z. G. 37.4 fs pulse generation in an Er: fiber laser at a 225 MHz repetition rate. *Opt. Lett.* **35**, 2858-2860 (2010).

33. Freitas, J. D. Recent developments in seismic seabed oil reservoir monitoring applications using fibre-optic sensing networks. *Meas. Sci. Technol.* **22**, 052001 (2011).

34. Cahill, J. P., Okusaga, O., Zhou, W., Menyuk, C. R. & Carter, G. M. Superlinear growth of Rayleigh scattering-induced intensity noise in single-mode fibers. *Opt.*

*Express* **23**, 6400–6407 (2015).

35. Chow, J. H., Littler, I. C. M., McClelland, D. E. & Gray, M. B. Laser frequency-noise-limited ultrahigh resolution remote fiber sensing. *Opt. Express* **14**, 4617-4624 (2006).

**Acknowledgements**

This work is supported in part by the Nature Science Foundation of China (Grant 61322509) and the National Research Foundation (NRF) of Korea (Grant 2012R1A2A2A01005544). X. Lu acknowledges the funding support from the Graduate School of Peking University. S. Zhang acknowledges the BK21+ Postdoctoral Fellowship from Ministry of Education, South Korea.

**Author contributions**

K. S. and J. K. developed the concept and led the project. X. L. and S. Z. set up the experiment, carried out the measurement, and obtained the final data. D. K. and C. J. set up the frequency comb source and assisted the experiment. X. C. and Z. Z. contributed to the noise improvement in the experiments. X. L., S. Z., J. K. and K. S. prepared the manuscript with inputs from all authors.

**Competing financial interests**

The authors declare no competing financial interests.

**Figure Legends**

**Figure 1.** TOF-based strain sensing detection principle: **a**. Synchronization of the microwave signal to the strain-free optical pulse train; **b**. Strain-influenced optical pulse train is phase-compared with the synchronized, referenced microwave signal to extract strain information.

**Figure 2.** Schematic of the experimental setup for TOF-based strain sensing. **a.** Local strain sensing experimental setup. **b.** Remote strain sensing experimental setup. FLOM-PD, fibre-loop optical-microwave phase detector; MLL, mode-locked laser; SMF, single-mode fibre; DCF, dispersion-compensation fibre; WDM, wavelength division multiplexer; OC, optical circulator; EDFA, Erbium-doped fibre amplifier; VCO, voltage-controlled oscillator.

**Figure 3.** Voltage-strain response of the TOF strain sensor. Voltage output from FLOM-PD2 is measured when a strain signal is applied to the strain sensor.

**Figure 4.** Measured strain power spectral density data over 7 decades [10 mHz – 100 kHz]. Curves **a** and **b** are the local and remote strain sensing resolutions of the present work, respectively. Curves **c** – **i** are from previous state-of-the-art results for comparison: **c.** Frequency comb-stabilized CW laser and FBG[9]; **d.** Free-running CW laser and FBG[9]; **e.** CW laser and FBG with dual PDH loops[5]; **f.** CW laser and Fabry-Perot sensor[11]; **g.** DFB fibre laser and a reference fibre laser locked to a FBG resonator[17]; **h.** CW laser locked to an atomic absorption line and FBG[16]; **i.** free-running CW laser and FBG[16]. Note that electric power supply pickup (peaks at 60 Hz and its harmonics) are indicated as dashed lines.

**Figure 1**

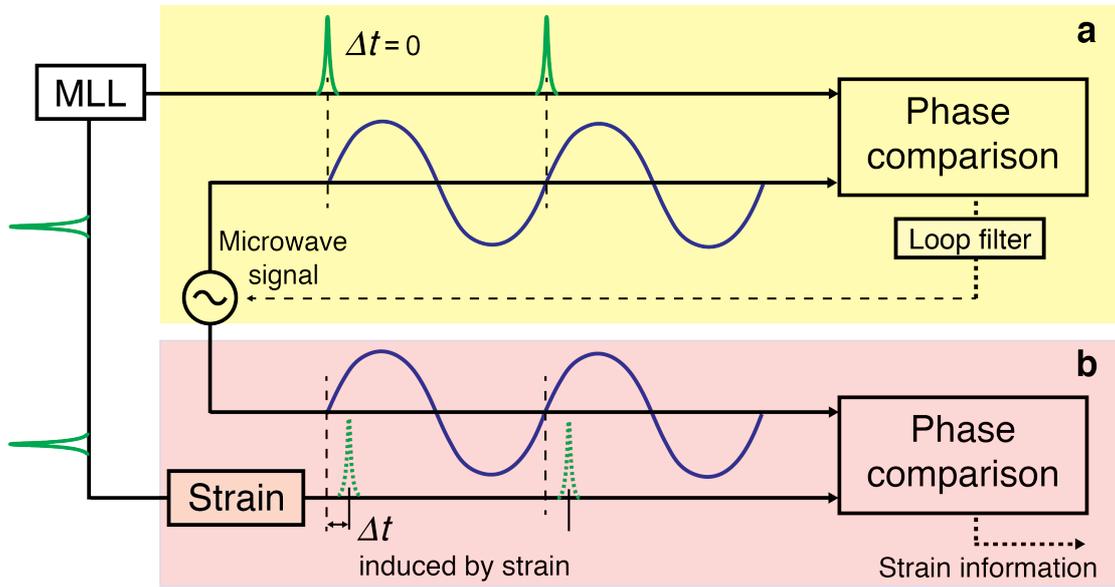

**Figure 2**

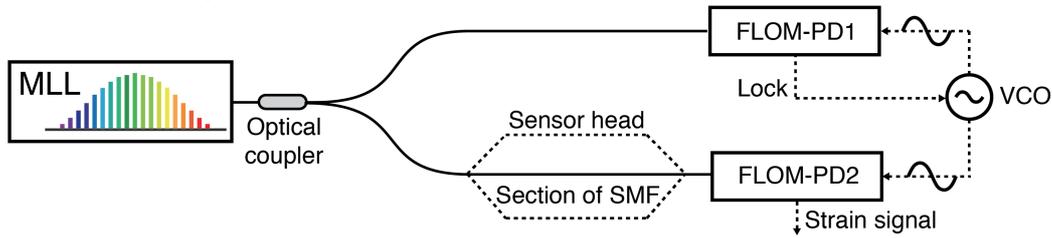

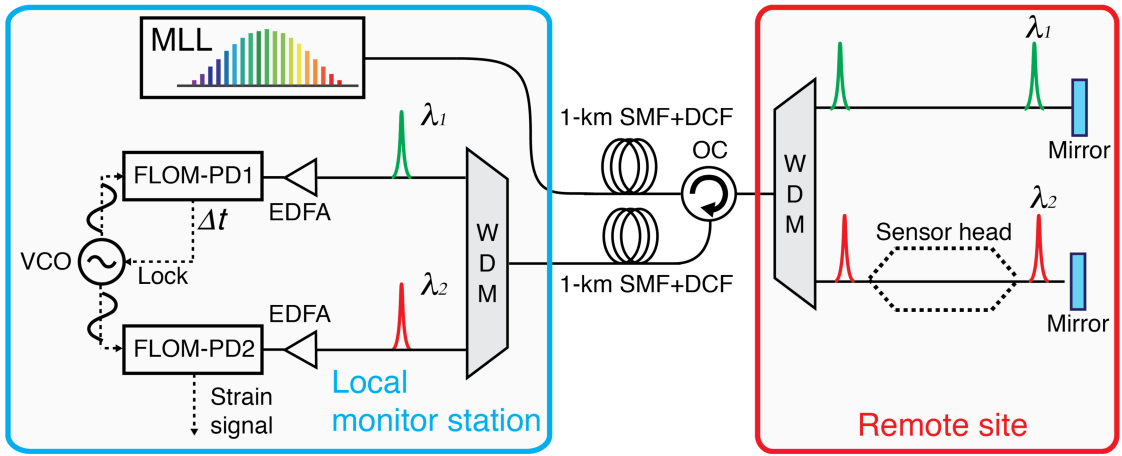

**Figure 3**

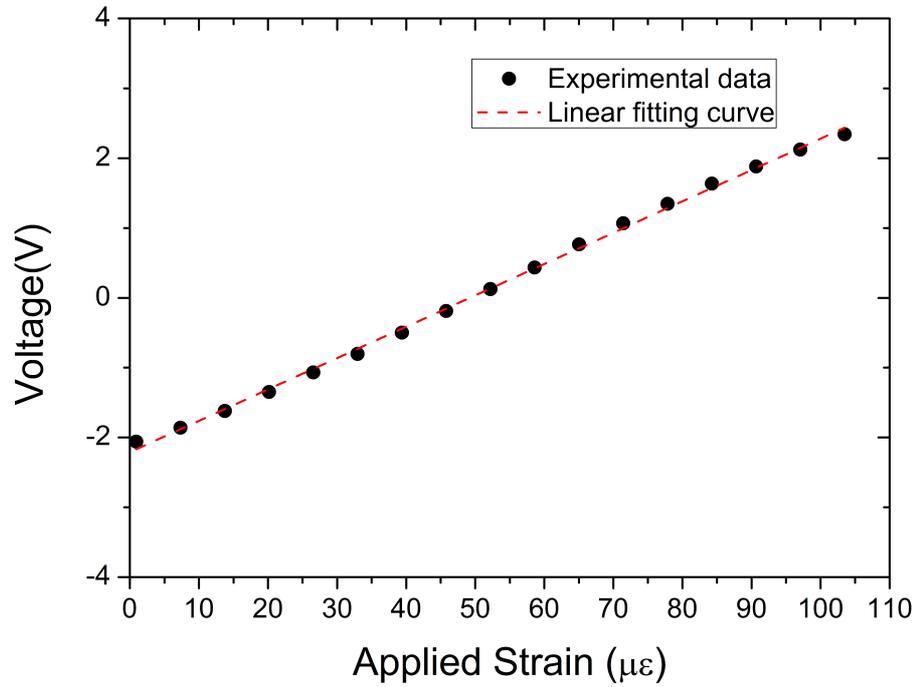

**Figure 4**

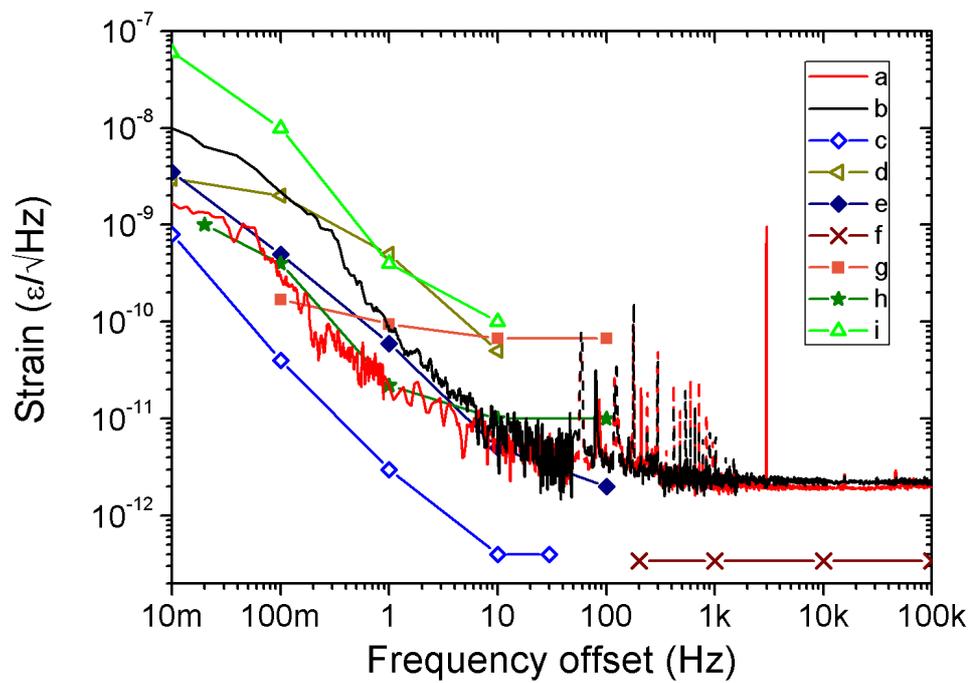